\documentclass[preprint,showpacs,showkeys,amsmath,amssymb,prd]{revtex4}
\usepackage{dcolumn}
\usepackage{epsfig}
\usepackage{bm}
\usepackage{rotating}
\usepackage{subfigure}
\usepackage{graphicx}
\usepackage{setspace}
\begin{document}
\begin{spacing}{2.0}

\title{
A Deduction from Dirac Sea  Model
}
\author{Jianguo~Bian$^1$}
\email{bianjg@mail.ihep.ac.cn}
\author{Jiahui~Wang$^2$}

\affiliation{\it
$^1$ Institute of High Energy Physics, Beijing 100049, China\\
$^2$ China Agricultural University,  Beijing 100083, China
}

\date{\today}

\begin{abstract}
Whether the Dirac sea model is right is verifiable.
Assuming the Dirac sea is a physical reality, we have one imagination that 
a negative energy particle(s) in the sea and a usual positive energy particle(s) 
will form a neutral atom.  The features of the atom can be studied using the nonrelativistic reduction of the Bethe-Salpeter equation, 
especially for a two body system. This work is dedicated to discuss the bound 
state of spin-0 and spin-$\frac{\displaystyle 1}{\displaystyle 2}$ constituents. 
The study shows that an atom consisting of a negative energy particle $(e^-$ or $hc)$  and a positive 
energy particle $(u\overline{d}$ or $hc)$ has some features of a neutrino, such as spin, lepton number and oscillation.  One deduction is that  
if the atom is a real particle, an atom $X$ consisting of two negative energy particles and a nucleus
 with double charges exists in nature, which can be observed 
in Strangeonium, Charmonium and Bottomonium decays
 $[s\overline{s}]$, $[c\overline{c}]$ and $[b\overline{b}]\rightarrow X \pi^-\pi^- e^+e^+~+~hc$ 
with $X$  invisible
at KLOE, CLEOc, BESIII, Belle and BaBar. 
\end{abstract}



\maketitle

 The Dirac sea$^{[1]}$ is a theoretical model of the vacuum as an infinite sea of particles with negative energy. It was first postulated by the British physicist Paul Dirac in 1930 to explain the anomalous negative-energy quantum 
states 
predicted by the Dirac equation, an extension of the Schr\"{o}dinger equation$^{[2]}$ for relativistic electrons. The positron, the antimatter counterpart of the electron, was originally conceived of as a hole in the Dirac sea, 
well before its 
experimental discovery in 1932. Despite its success, the idea of the Dirac sea tends not to strike people as very elegant. The development of quantum of filed in the 1930s reformulate the Dirac equation in a way that treats the 
positron as a real particles rather than the absence of a particle and make the vacuum the  state in which no particles exist instead of an infinite sea of particles.  
Therefore a question appears whether the idea of the Dirac sea is completely wrong and how to verify it?
For instance, for the process $\gamma\rightarrow e^+e^-$, 
the description of the quantum of filed is that a gamma converts into a pair of electron and positron,
but the language of the Dirac sea model is that a gamma hits a nagetive energy electron; the electron excites into an positive energy state and an positron appears simutaniously
as a "hole". The two pictures are quite different. Either of the two is right?
The aim of this article is  to propose an observation scheme to verify the correction of the Dirac sea.
We assume the idea of the Dirac sea is right, extend it to including negative energy positrons and the other leptons
and use the nonrelativistic reduction of the Bethe-Salpeter equation$^{[3]}$
to investigate a possibility that a usual positive energy particle $(u\overline{d}$ or $hc)$   and a negative energy particle $(e^-$ or $hc)$  in the Dirac sea  will form a neutral atom 
and study special properties of this kind of atoms.
The article is organized as follow. The  nonrelativistic reduction of the Bethe-Salpeter equation about a two body system of spin-0 and spin-$\frac{\displaystyle 1}{\displaystyle 2}$
constituents is presented in paragraphs 2.
The detailed steps to solve the equation with $E<0$ and  $E>0$ are described in paragraphs 3 and 4, respectively.
The physical information is extracted from the solution to the equation in paragraph 5.
The final paragraph arrives at a summary.

\begin{figure}[hbtp]
  \begin{center}
    \epsfig{figure=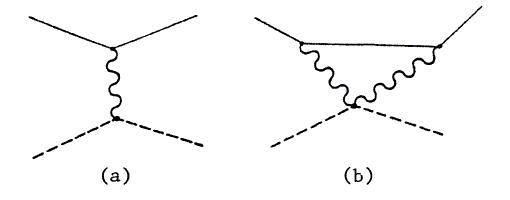, width=6.5cm,  height=3.5cm}
    \caption{Feynman diagrams for (a) single-photon exchange and (b) the
"seagull" interaction. The solid, dashed and wavy lines represent  a
spin-$\frac{\displaystyle 1}{\displaystyle 2}$ fermion, a spin-0 boson and
a photon, respectively.  }
    \label{labelcoulombseagull}
  \end{center}
\end{figure}

The nonrelativistic reduction of the Bethe-Salpeter equation for a charge-anticharge two body system of spin-0 and spin-$\frac{\displaystyle 1}{\displaystyle 2}$ 
positive energy constituents ($E_{s}=\sqrt{q_s^2+M_s^2}$, $E^\prime _{f}=\sqrt{q_s^2+m_f^2}$) through the single-photon and "seagull" interaction shown
in Fig. (\ref{labelcoulombseagull}) has been given in Ref.${[4]}$,
\begin{eqnarray}
E^\prime\psi({\bf r})=-(\frac{\displaystyle 1}{\displaystyle 2M_s}+\frac{\displaystyle 1}{\displaystyle 2m_f})\nabla^2\psi({\bf r})-\frac{\displaystyle q^2}{\displaystyle r}\psi({\bf r})
-\frac{\displaystyle 7q^4}{\displaystyle 4M_s}\frac{\displaystyle 1}{\displaystyle r^2}\psi({\bf r}),
\label{labelsb1}
\end{eqnarray}
where $M_s$ and $q$ are the mass and charge of the spin-0 particle and $m_f$ and $-q$ are the mass and charge of the spin-$\frac{\displaystyle 1}{\displaystyle 2}$ particle,
the nonrelativistic energy $E^\prime$ is related to the relativistic energy $E^\prime_r$ by $E^\prime_r=M_s+m_f+E^\prime.$
From Eq. (\ref{labelsb1}) it is straightforward to make  the nonrelativistic reduction of the Bethe-Salpeter equation for a charge-anticharge two body system in which the spin-0 constituent
is a positive energy particle while the  spin-$\frac{\displaystyle 1}{\displaystyle 2}$ constituent is a negative energy particle $E_f=-\sqrt{q_f^2+m_f^2}$,
\begin{eqnarray}
E\psi({\bf r})=-(\frac{\displaystyle 1}{\displaystyle 2M_s}-\frac{\displaystyle 1}{\displaystyle 2m_f})\nabla^2\psi({\bf r})-\frac{\displaystyle q^2}{\displaystyle r}\psi({\bf r})
-\frac{\displaystyle 7q^4}{\displaystyle 4M_s}\frac{\displaystyle 1}{\displaystyle r^2}\psi({\bf r}),
\label{labelsb2}
\end{eqnarray}  
where the nonrelativistic energy $E$ is related to the relativistic energy $E_r$ by
\begin{eqnarray} 
E_r=M_s-m_f+E\geq0,
\label{labelenergy1}
\end{eqnarray}
It is worth emphasizing that in Eq. (\ref{labelsb2}), $M_s$ is a physical state, while $m_f$ is a vacuum state.
One considers that a vacuum state can not be observed as a physical state directly, 
but there are interactions between physical states and vacuum states.
In Eq. (\ref{labelenergy1}), $-m_f$ alone does not represent a negative mass particle, instead $-m_f$ and $-\frac{\displaystyle q_f^2}{\displaystyle 2m_f}$ together
describe the nonrelativistic reduction of the relativistic energy of a negative energy particle $E_f=-\sqrt{q_f^2+m_f^2}$.
In spherical polar coordinates 
\begin{eqnarray}
\psi({\bf r})=R(r)\Theta(\theta)\Phi(\phi).
\label{labelsb3}
\end{eqnarray}
In this work, only radial wave function $R(r)$ is worth discussing,
\begin{eqnarray}
\frac{\displaystyle 1}{\displaystyle r^2}\frac{\displaystyle d}{\displaystyle dr}{\Big (}r^2\frac{\displaystyle dR}{\displaystyle dr}{\Big)}-{\Big [}\frac{\displaystyle 2\mu}{\displaystyle \hbar^2}
{\Big (}E+\frac{\displaystyle q^2}{\displaystyle r}
+\frac{\displaystyle 7q^4}{\displaystyle 4M_s}\frac{\displaystyle 1}{\displaystyle r^2}{\Big )}
+\frac{\displaystyle l(l+1)}{\displaystyle r^2}{\Big ]}R=0.
\label{labelsb4}
\end{eqnarray}
Let $R(r)=\frac{\displaystyle u(r)}{\displaystyle r}$, one has
\begin{eqnarray}
\frac{\displaystyle d^2 u}{\displaystyle dr^2}-{\Big [}\frac{\displaystyle 2\mu}{\displaystyle \hbar^2}
{\Big (}E+\frac{\displaystyle q^2}{\displaystyle r}
+\frac{\displaystyle 7q^4}{\displaystyle 4M_s}\frac{\displaystyle 1}{\displaystyle r^2}
{\Big )}+\frac{\displaystyle l(l+1)}{\displaystyle r^2}{\Big ]}u=0,
\label{labelsb5}
\end{eqnarray}
where the reduced mass $\mu=\frac{\displaystyle m_fM_s}{\displaystyle M_s-m_f}.$

For $E<0$,  one has
\begin{eqnarray}
\frac{\displaystyle d^2 u}{\displaystyle dr^2}+{\Big [}\frac{\displaystyle 2\mu}{\displaystyle \hbar^2}
{\Big (}\mid E \mid-\frac{\displaystyle q^2}{\displaystyle r}
-\frac{\displaystyle 7q^4}{\displaystyle 4M_s}\frac{\displaystyle 1}{\displaystyle r^2}
{\Big )}-\frac{\displaystyle l(l+1)}{\displaystyle r^2}{\Big ]} u=0. 
\label{labelsb6}
\end{eqnarray}
Eq. (\ref{labelsb6}) has the same form of the Schr\"{o}dinger equation for two body scattering in repulsive fields$^{[5]}$
\begin{eqnarray}
\frac{\displaystyle d^2 u}{\displaystyle dr^2}+{\Big [}\frac{\displaystyle 2\mu^\prime}{\displaystyle \hbar^2}
{\Big (} E -\frac{\displaystyle Zq^2}{\displaystyle r}{\Big )}-\frac{\displaystyle l(l+1)}{\displaystyle r^2}{\Big ]} u=0,  
\label{labelsb7}
\end{eqnarray}
where $E>0$ and the reduced mass $\mu^\prime=mM/(M+m).$
Let $\alpha={\Big(} \frac{\displaystyle 8\mu\mid E\mid}{\displaystyle \hbar^2} {\Big)}^\frac{\displaystyle 1}{\displaystyle 2},~\eta=\frac{\displaystyle q^2}{\displaystyle \hbar}{\Big (}
\frac{\displaystyle \mu}{\displaystyle 2 \mid E \mid} {\Big )}^\frac{\displaystyle 1}{\displaystyle 2}$ and $~\rho=\frac{\displaystyle \alpha r}{\displaystyle 2},
~\gamma=\frac{\displaystyle 7q^4}{\displaystyle \hbar^2}\frac{\displaystyle \mu}{\displaystyle 2M_s},$
Eq. (\ref{labelsb6}) becomes
$$
\frac{\displaystyle d^2 u}{\displaystyle d\rho^2}+{\Big [}1-\frac{\displaystyle 2 \eta}{\displaystyle \rho}
-\frac{\displaystyle \gamma}{\displaystyle \rho^2}
-\frac{\displaystyle l(l+1)}{\displaystyle \rho^2}{\Big ]}u=0,
$$
i.e.
\begin{eqnarray}
\frac{\displaystyle d^2 u}{\displaystyle d\rho^2}+{\Big [}1-\frac{\displaystyle 2 \eta}{\displaystyle \rho}
+\frac{\displaystyle 
\frac{\displaystyle 1}{\displaystyle 4}-L^2}{\displaystyle \rho^2}{\Big ]}u=0, 
\label{labelsb8}
\end{eqnarray}
where 
\begin{eqnarray}
L=\sqrt{(l+\frac{\displaystyle 1}{\displaystyle 2})^2+\gamma}.
\label{labelbigl}
\end{eqnarray}
Eq. (\ref{labelsb8}) is called the Coulomb wave equation$^{[5]}$, which has two independent solutions
\begin{eqnarray}
u(\eta,\rho)=C_1F_L(\eta,\rho)+C_2G_L(\eta,\rho)~~~(C_1,~C_2~{\rm constants}),
\end{eqnarray}
where $F_L(\eta,\rho)$ is the regular Coulomb wave function and $G_L(\eta,\rho)$ is the irregular Coulomb wave function.
The asymptotic forms of the Coulomb wave functions can be obtained from the theory of the confluent hypergeometric function.
\begin{eqnarray}
F_L(\eta,\rho)=C_L(\eta)\rho^{L+1} [1+O(\rho) ]~~~{\rm as}~\rho\rightarrow 0. \label{label4}
\end{eqnarray}
\begin{eqnarray}
F_L(\eta,\rho)=sin(\rho-\eta ln(2\rho)-\frac{\displaystyle 1}{\displaystyle 2}L\pi+\sigma_L(\eta))+O(\rho^{-1})~~~{\rm as}~\rho\rightarrow \infty, \label{label5}
\end{eqnarray}
where $\sigma_L(\eta)={\rm arg}\Gamma(L+1+i\eta)$.
\begin{eqnarray}
G_L(\eta,\rho)=O(\rho^{-L})~~~{\rm as}~\rho\rightarrow 0. 
\label{label6}
\end{eqnarray}
\begin{eqnarray}
G_L(\eta,\rho)=cos(\rho-\eta ln(2\rho)-\frac{\displaystyle 1}{\displaystyle 2}L\pi+\sigma_L(\eta))+O(\rho^{-1})~~~{\rm as}~\rho\rightarrow \infty. \label{label7}
\end{eqnarray}
From Eqs. (\ref{labelbigl},~\ref{label6}), it follows that $\int_0^\infty r^2 \frac{\displaystyle G_L^2(\eta,r)}{\displaystyle r^2} 
 dr$ is not integrable for any $l$.
From Eqs.(\ref{label5},~\ref{label7}), it is noted that $F_L(\eta,\rho)$ and $G_L(\eta,\rho)$ describe the scattered particle radial wave functions of Eq. (\ref{labelsb5}).

For $E>0$, Eq. (\ref{labelsb5}) becomes
\begin{eqnarray}
\frac{\displaystyle d^2 u}{\displaystyle d\rho^2}+{\Big [}-1-\frac{\displaystyle 2 \eta}{\displaystyle \rho}
+\frac{\displaystyle
\frac{\displaystyle 1}{\displaystyle 4}-L^2}{\displaystyle \rho^2}{\Big ]}u=0.
\label{labelsb9}
\end{eqnarray} 
Let
\begin{eqnarray}
u=e^{\displaystyle -\rho}f(\rho),
\label{labelsb10}
\end{eqnarray}
$f(\rho)$ satisfies
\begin{eqnarray}
f^{\prime\!\prime}(\rho)-2f^{\prime}(\rho)+
{\Big [}
-\frac{\displaystyle 2\eta}{\displaystyle \rho}
+\frac{\displaystyle
\frac{\displaystyle 1}{\displaystyle 4}-L^2}{\displaystyle \rho^2}
{\Big ]}f(\rho)=0.
\label{labelsb10b}
\end{eqnarray}
Let
\begin{eqnarray}
f(\rho)=\sum\limits_{n=0}^{\infty}a_n \rho^{n+s}
\label{labelfrho}
\end{eqnarray}
and substitute it to eq. (\ref{labelsb10}), one has the relation
\begin{eqnarray}
a_{n+1}=\frac{\displaystyle 2\eta+2(n+s)}{\displaystyle (n+1+s)(n+s)-\gamma-l(l+1)}a_n.
\label{labelsb11}
\end{eqnarray}
Due to $a_0\neq 0$, one has
\begin{eqnarray}
s(s-1)=l(l+1)+\gamma.
\label{labelsb12}
\end{eqnarray}
The s solution is 
\begin{eqnarray}
s=\frac{\displaystyle 1}{\displaystyle 2}\pm \sqrt{(l+\frac{\displaystyle 1}{\displaystyle 2})^2+\gamma}.
\label{labelsb12b}
\end{eqnarray}
From eqs. (\ref{labelfrho},~\ref{labelsb11}), it is required
\begin{eqnarray}
2s>-1.
\label{labelsb13}
 \end{eqnarray}
and
\begin{eqnarray}
\eta+n+s\leq 0,
\label{labelsb14} 
 \end{eqnarray}
i.e.
\begin{eqnarray}
s\leq 0,
\label{labelsb15}
 \end{eqnarray}
due to $\eta\geq 0$ and $n\geq 0$
so that $\int_0^\infty r^2R^2(r)dr$ is integrable. 
From eq. (\ref{labelsb13}), one has
\begin{eqnarray}
l(l+1)+\gamma<\frac{\displaystyle 3}{\displaystyle 4}.
\label{labelsb16}
\end{eqnarray}
From  eq. (\ref{labelsb15}), one has
\begin{eqnarray}
l(l+1)+\gamma\geq0.
\label{labelsb17}
\end{eqnarray}
Combining eqs. (\ref{labelsb16},~\ref{labelsb17}), one has
\begin{eqnarray}
l=0,
\label{labelsb18}
\end{eqnarray}
and an exact solution of $f(\rho)$
\begin{eqnarray} 
s=\frac{\displaystyle 1}{\displaystyle 2}-\sqrt{\frac{\displaystyle 1}{\displaystyle 4}+\gamma},
\label{labelsb18b}
\end{eqnarray}  
\begin{eqnarray} 
\eta=-s=\sqrt{\frac{\displaystyle 1}{\displaystyle 4}+\gamma}-\frac{\displaystyle 1}{\displaystyle 2},
\label{labelsb19}
\end{eqnarray}  
\begin{eqnarray}
a_{n}=0,~n=1,~2,~...
\label{labelsb20}
\end{eqnarray}
It is easy to check if the spin-0 constituent is a $\pi^\pm$ and the  spin-$\frac{\displaystyle 1}{\displaystyle 2}$ is an 
$e^{\mp}$ or $\mu^{\mp}$, $0<\gamma<\frac{\displaystyle 3}{\displaystyle 4}$.

Now, we turn to extract the physical information from  above discussion.
Eq. (\ref{labelsb5}) has a bound state solution described by Eqs. (\ref{labelsb18}-\ref{labelsb20}) for the  energy  
$E=\frac{\displaystyle q^2}{\displaystyle \hbar}\frac{\displaystyle \mu}{\displaystyle 2\eta^2}$ and only the 
orbital quantum number $l=0$. Eq. (\ref{labelsb5}) also has a scattering state solution for the continuum $-M_s+m_f<E<0$, here the lower limit is given by Eq. (\ref{labelenergy1}).
Therefore the bound state with  energy  $E>0$ will transit to the scattering state with  energy  $E<0$, then the positive energy constituent and another negative energy particle in the Dirac sea
will form a new bound state and the new bound state will transit. The process of formation and transition cycles at intervals of 
\begin{eqnarray}
\Delta t\geq \frac{\displaystyle \hbar}{\displaystyle 2{\Big (}\frac{\displaystyle q^2}{\displaystyle \hbar}\frac{\displaystyle \mu}{\displaystyle 2\eta^2}+M_s-m_f{\Big )}}
\end{eqnarray}
based on the uncertainty principle 
$\Delta t \Delta E\geq\frac{\displaystyle \hbar}{\displaystyle 2}$.
 If the positive energy particle consists of $u\overline{d}$ and the negative energy particle is an electron, 
 $u\overline{d}$ and $e^-$ will form an atom we call $X(e^-u\overline{d})$. 
The atom has some features of the neutrino  $\nu_e$. 
The spin quantum is  $\frac{\displaystyle 1}{\displaystyle 2}$ . 
The lepton number is 1. $X(e^-u\overline{d})$ will transit to the scattering state
and then a new atom  $X(e^-u\overline{d})$ or $X(\mu^-u\overline{d})$  will form as a neutrino oscillates.
However, in this simple work, the mass of the bound state can not be obtained due to the non-relativistic reduction.
If $X(e^-)$ is a real particle,  one deduction is that four quarks $u\overline{d}u\overline{d}$ and two negative energy electrons $e^-e^-$ can also form
an  atom we call $X(2(e^-u\overline{d}))$.
The $X(2(e^-u\overline{d}))$ can be observed in
Strangeonium, Charmonium and Bottomonium decays $[s\overline{s}]$, $[c\overline{c}]$ and $[b\overline{b}]\rightarrow X(2(e^-u\overline{d})) \pi^-\pi^- e^+e^+~+~hc$ with $X(2(e^-u\overline{d}))$ and $hc$
invisible at KLOE$^{[6]}$,  CLEOc$^{[7}$, BESIII$^{[8]}$, Belle$^{[9]}$ and BaBar$^{[10]}$.

In summary, the Dirac sea is verifiable. If the Dirac sea is a physics reality, the process $\gamma\rightarrow e^+e^-$ cannot be lonely observable.
This work shows that
$u\overline{d}$ and a negative energy electron $e^-$ will form an atom $X(e^-u\overline{d})$ with the lepton number equal to 1 and
$u\overline{d}u\overline{d}$ and two  negative energy electrons $e^-e^-$ will form another atom $X(2(e^-u\overline{d}))$ with the lepton number equal to 2.
The former is  hard to distinguish from a neutrino while
the latter can be observed exclusively.
One should pay attention to that in contrast to those well-known normal atoms which consist of all positive energy components and have long life-time,  
this kind of atoms are unstable for they oscillate from a state which has electron components to another state which has muon components. 
If the $X(2(e^-u\overline{d}))$ is not observed, we should doubt the correction  of the Dirac sea.

\newpage

\end{spacing}

\end{document}